# The Great Rift in Physics:

## The Tension Between Relativity and Quantum Theory

Tim Maudlin

New York University

John Bell Institute for the Foundations of Physics

Abstract: It is commonly remarked that contemporary physics faces a challenge in reconciling quantum theory with Relativity, specifically General Relativity as a theory of gravity. But "challenge" is too mild a descriptor. Once one understands both what John Bell proved and what Einstein himself demanded of Relativity it becomes clear that the predictions of quantum theory—predictions that have been verified in the lab—are flatly *incompatible* with what Einstein wanted and built into General Relativity. There is not merely a tension but an incompatibility between the predictions of quantum theory and Relativity, and what has to give way is the Relativistic account of space-time structure and dynamics.

It is commonly said that modern physics rests on two great pillars: Quantum Theory and the General Theory of Relativity. Lately, a third pillar is sometime offered, namely Thermodynamics or more generally Statistical Physics. But statistical physics in general, and thermodynamics in particular, do not postulate anything new or novel at the level of fundamental ontology or fundamental dynamics. Statistical techniques concern ways of treating systems with many, many fundamental degrees of freedom without having to take precise account of the values of each of them. Statistical techniques are methodologically and pragmatically important, and also offer a type of understanding not afforded by only following the precise trajectory of the exact microstate of a system. So in the sequel we will leave consideration of statistical matters aside and focus on fundamental ontology and dynamics alone. Quantum Theory and Relativity each contributes to our best understanding of fundamental physics in terms of ontology and dynamics.

However, it is also universally acknowledged that quantum theory and Relativity—while each has achieved outstanding empirical success in a certain delimited range of cases—"do not play together well". At the most generic level, one may say that since there is only one physical world, the physics that governs it must ultimately be expressible as a single unified physical theory from which each of quantum theory and Relativity "emerge" as a special case in specific circumstances (for example when the gravitational mass in a particular situation is alternatively the dominant physically important feature or instead a quite minor and negligible feature). That is, one expects to be able to see from the perspective of this single unified theory why quantum theory alone works as well as it does where it does, and why General Relativity works as well as it does where it does, and also what should be expected in situations where both gravitational and quantum mechanical effects are substantial and neither can be ignored (for example in some situations in astrophysics).

The main undertaking of this essay is to consider in detail just what is meant by saying that the two theories do not play together well, or that there is some difficulty "putting them together" into one, single, unified physical theory. I think that the depth and intractability of that difficulty has not been sufficiently appreciated, and that progress on this question will not be forthcoming until it is.

What sorts of problems could arise trying to put two theories together in this way, so each emerges from the resulting theory under specific conditions? One thing that is sometimes said is that quantum theory and Relativity employ different mathematical formalisms, so it is not obvious even how to begin trying to unite them. For example, the mathematics of standard quantum theory is based on a Lagrangian or Hamiltonian formalism that in turn was developed in the context of Newtonian mechanics. One avenue of research along this direction starts by trying to express General Relativity in Hamiltonian form, and then applying a "standard quantization procedure". The result is the Wheeler-Dewitt equation, which exhibits some worrying features. The main one is this: in standard quantum theory, the Hamiltonian operator

(which results from taking the classical Hamiltonian *function* of Newtonian theory and converting into an operator by "putting the hats on") does not do what it does in standard quantum theory, namely generate the time evolution of the quantum state. Rather, according to the Wheeler-DeWitt equation, the Hamiltonian operator *annihilates* the quantum state of the universe, that is, operating with the operator on the quantum state yields the zero vector.

In normal circumstances, such a result would be regarded as an indicator of what Imre Lakatos called a "degenerating research programme", and some alternative to the Wheeler-DeWitt approach would be sought. In the rather extraordinary circumstances that theoretical physics has managed to get itself into, the result has rather just been denominated "the problem of time", and it has been seriously contended that in fact there is no time evolution of the physical universe, and nothing really changes, but all appearance of temporal change is just some sort of illusion. (As aficionados of Parmenides will recognize, there is really nothing new under the sun.)

Much could be said about this particular approach to "quantum gravity", but as extreme as it is, the details of the Wheeler-DeWitt equation as a basis for quantum gravity will not be our focus here.

The present point is rather this: if the fundamental difficulty reconciling or combining quantum theory (the theory of the electromagnetic, Weak, and Strong nuclear forces) with General Relativity (the theory of gravitation) were just a matter of mathematical compatibility, it might admit of a purely mathematical solution. If that were all that stood in the way of uniting the two pillars into one coherent theory, then it would be possible to wake up one morning to the news that some clever mathematician had just solved the problem, perhaps by reframing one or both of the theories in some novel mathematical framework. Technical problems admit of technical solutions. But my aim here is to argue that the conflict or tension between General Relativity and Quantum Theory runs much deeper than that.

On the subject of mathematical structure, it is worthwhile to pause a moment over the natural mathematical language in which General Relativity is presented (unlike the language of Hamiltonian or Lagrangian dynamics, into which it must be massaged or even forced). General Relativity, on its own, is usually presented in the mathematical language of differential geometry.

That is the sub-branch of geometry that deals with geometrical structure *locally* rather than *globally*. In differential geometry, one need not specify the global geometrical structure of a space entirely all at once, but rather can subdivide the space into arbitrarily many arbitrarily small open sets, called "charts", specifying the geometry of each chart (nearly) independently of all the rest. The "nearly" in the foregoing is connected to the requirement that all the charts be *open* sets: that means that every chart must to some degree overlap on its edges with each neighboring open chart, and the two accounts of the geometry much coincide in the joint region of overlap. But once one has specified the geometry of each chart in an "atlas", where the union of all the charts in the atlas contains the entire space, the entire geometry is thereby fixed. In that sense, using differential geometry means that one can treat the geometrical structure of a space as just one little piece after another (with tiny overlaps of the pieces).

This particular feature of differential geometry expresses an important idea about the direction that physics progressed from Newton until the early 20$^{th}$ century. In a famous letter to Born (and importantly setting aside quantum theory!) Einstein wrote:

> If one asks what, irrespective of quantum mechanics, is characteristic of the world of ideas of physics, one is first of all struck by the following: the concepts of physics relate to a real outside world, that is, ideas are established relating to things such as bodies, fields, etc., which claim "real existence" that is independent of the perceiving subject—ideas which, on the other hand, have been brought into as secure a relationship as possible with the sense data. It is further characteristic of these physical objects that they are thought of as arranged in a space-time continuum. An essential aspect of this arrangement of things in physics is that they lay claim, at a certain time, to an existence independent of one another, provided these objects "are situated in different parts of space". Unless one makes this kind of assumption about the independence of the existence (the "being thus") of objects which are far apart from one another in space—which stems in the first place from everyday thinking—physical thinking in the familiar sense would not be possible….This principle has been carried to extremes in the field theory by localizing the elementary objects on which it is based and which exist

> independently of each other, as well as the elementary laws which have been postulated for it, in the infinitely small (four-dimensional) elements of space.[1]

Just as differential geometry describes the geometrical structure of a space arbitrarily-small-piece by arbitrarily-small-piece, each part independently of the others (except in the overlaps of neighboring regions), so Einstein saw the progress of physics as extending that sort of *locality* to both the *material ontology* and the *dynamical laws* postulated by physics. The values of classical fields (such as the electric and magnetic fields of Maxwell) can similarly be specified tiny patch by tiny patch in space or in space-time, requiring only agreement among the patches where they overlap. And, even more importantly for Einstein, the *fundamental dynamical laws* of classical field theory (again, such as Maxwellian electrodynamics) displayed exactly the same feature. The *laws themselves* could be formulated in terms of local equations. So determining whether the electric and magnetic fields obey Maxwell's equations universally can be reduced to the question of whether they are satisfied individually in each small open set into which the space-time has been (arbitrarily) divided.

Not every possible physical law has this feature of locality. Let's consider an example of a possible theory in which it fails.

It is often said (although it is not true) that Newton's gravitational theory postulated some gravitational action-at-a-distance: one massive body could have a gravitational effect on another massive body *instantaneously* and even if there were a physical vacuum between them, so the effect was not mediated by any continuous causal process. Newton said no such thing in *Principia*. Indeed, it was exactly at the point of inquiring *what the ultimate cause of the universal gravitational force is* that Newton famously declared "Hypothesis non fingo" ("I frame no hypothesis"). Newton's contention (which was not true by a modern understanding of "deduce") was that the *Principia* would avoid all contentious and questionable guesses and confine itself only to *what can be deduced from the phenomena*. In the modern logical sense of "deduce", such a feat is plainly impossible. As we say, theory is always underdetermined by the phenomena or the evidence: the same evidence can, in principle, be "saved by different theories". If so, then the correct theory cannot be deduced from the phenomena in a certainly truth-preserving way. But Newton did not even attempt to present some *purely logical* derivation of his theory of gravity from the phenomena. For example, his arguments all rely on the postulation of his three laws of motion, and he never purports to have derived those laws themselves from the phenomena. Newton's method is more complicated—and more vulnerable to error—than pure logical deduction.

---
[1] Born, Max, *The Born-Einstein Letters*, trans. I. Born, Walker, 1971, pp. 170-1.

It would be closer to the mark to say that in Newton's presentation—once the fundamental laws of motion are granted—the exact form of the gravitational force is *directly motivated* from consideration of observable phenomena, although not *logically deduced* from them. In particular, none of the phenomena Newton considers—falling apples on the surface of the Earth and the orbits of the planets and their moons—suggest that gravity involves any sort of instantaneous distant action. The situations as very nearly stationary as far as gravity is concerned: the motion of the Earth around the Sun, for example, is extremely slow and the motion of the Sun itself—according to the theory—would be quite small. At the level of precision that the phenomena were recorded, one could not possibly distinguish between various "speeds of gravitational effect". For example, even if the gravitational influence of the Sun took an hour to reach the Earth (in fact it is about 8 minutes and 19 seconds), the direction and degree of gravitational force on the Earth could not be observationally distinguished from an instantaneous one by the data Newton had. And what we know for sure is that Newton himself rejected any instantaneous, unmediated forces between bodies. In his letter to Richard Bentley of February 25, 1692 Newton writes:

> It is inconceivable that inanimate brute matter should, without the mediation of something else, which is not material, operate upon, and affect other matter without mutual contact; as it must do, if gravitation, in the sense of Epicurus, be essential and inherent in it. And this is the reason why I desired you would not ascribe innate gravity to me. That gravity should be innate, inherent, and essential to matter, so that one body may act upon another at a distance through a vacuum, without the mediation of anything else, by and through which their action and force may be conveyed from one to another, is to me so great an absurdity, that I believe no man who has in philosophical matters a competent faculty of thinking, can ever fall into it. Gravity must be caused by an agent acting constantly according to certain laws; but whether this agent be material or immaterial, I have left to the consideration of my readers.[2]

Any reader of Newton, then, who postulates that the gravitational interaction between the Sun and the Earth is mediated by (say) particles of some sort, which travel at some finite speed, will not find anything in *Principia* to oppose that hypothesis. And if the force were mediated by such particles, then the ultimate theory could be local in Einstein's sense (most obviously if they propagate along light cones).

But we were after a theory that would *violate* Einstein's observation. An action-at-a-distance-without-any-mediation gravitational theory (whether instantaneous or not, but most clearly if instantaneous) would clearly violate the condition. Suppose, for example, body A is attracted gravitationally by distant body B, and body B is subjected to a sudden change of position, having been jerked by a cord. Then the gravitational acceleration of body A will also suddenly change, despite there being no alteration in the *local* physical conditions. So Einstein's condition would fail: given complete information about the behavior of matter in a small four-dimensional volume including the sudden change in the acceleration of A, one could not determine whether the laws of nature had been violated. Only investigation of what happened to B would account for the behavior of A, and B need not be in the small four-dimensional volume under consideration. On the other hand, if B could only influence A via some particles or fields that propagate continuously between them, then the change in those particles or fields would show up in the small volume, and one could check whether the laws hold there.

In sum, the sort of locality of both the ontology and the laws that Einstein describes is a substantial constraint on any physical theory. The constraint becomes even stronger if one demands that the mediating particles or fields not propagate faster than light. If in addition to such a restriction on causal influences (which can be stated in a completely Relativistic language, since one need only refer to the light-cone structure of a space-time) one also demands no temporal action-at-a-distance (that is, that the dynamics is a Markov process), one arrives at the condition of Relativistic causal locality or Bell locality. It is precisely this condition that Bell assumes in the proof of his theorem, and that is assumed in related theorems such as the CHSH inequality.

---

[2] Cohen, I. Bernard, ed. 1978. *Isaac Newton's Papers & Letters on Natural Philosophy and Related Documents*. Reprint 2014 ed. edition. Harvard University Press.

To repeat again: classical Maxwellian electromagnetic theory and the General Theory of Relativity both respect Bell locality, so neither of those theories could predict or explain violations of Bell's Inequality (or the CHSH Inequality, or the three-party correlations derived in the GHZ example, as we will see). For Einstein's purposes, namely constructing theories that could explain the successes of Maxwell's electromagnetics and Newtonian gravitation, the Relativistic space-time structure (and the causal locality formulable in terms of it) sufficed. As far as Einstein was concerned, the space-times of Special and General Relativity served the purposes for which he constructed them.

But those purposes did not include recovering all the predictions of *quantum* theory.

Einstein, as is well known, was never satisfied with quantum theory as it was exposited by Bohr and the Copenhagen school. He saw immediately—as soon as the "new quantum theory" was formulated—that Bohr and company were committing themselves to "spooky action-at-a-distance" or "telepathy"—whether they acknowledged it or not. Einstein's early arguments to that conclusion did not reference violations of Bell's Inequality, of course, or even require consideration of systems with more than one particle. A single radioactive atom (of the sort that played the role of possible triggering mechanism in Schrödinger's cat set-up) sufficed to make the point. The quantum-mechanical wavefunction ascribed to the single atom would exhibit spherical symmetry: it would, as it were, constantly "leak out" from the atom equally in all directions, forming spherical waves. Einstein imagined surrounding such a system with a spherical detector, so as not to break the symmetry. Then the initial wavefunction of the atom-and-detector system would be spherically symmetric, and pure Schrödinger evolution of that wavefunction could not break the symmetry. That much is easy to see.

Nonetheless, in any such concrete system in a laboratory the spherical symmetry would eventually be broken. A single spot would form somewhere on the spherical detector. Since (by

Curie's principle) no deterministic evolution from a symmetric initial condition can yield a non-symmetric final condition, the symmetry has to be broken in some way that goes beyond Schrödinger evolution of the wavefunction. And there are only two possible ways to break it: in the initial condition itself, or in the dynamical evolution which cannot (therefore) be deterministic.

Breaking the symmetry in the initial condition—asserting that the spot formed at one place of the spherical detector rather than another because there was already a corresponding asymmetry in the system *ab initio*—clearly entails denying that the wavefunction of the system provides a complete physical description of it. For the initial wavefunction is spherically symmetric, so whatever initially breaks the symmetry is a physical condition not reflected in the wavefunction. Since Bohr's approach demanded that the wavefunction be complete (no "hidden variables"), that option was not available to them.

The only other escape available, then, was to deny that the dynamics is always deterministic, e.g. in accord with Schrödinger's equation. And that is precisely what Bohr and Heisenberg and (later) von Neumann did. They insisted that the physical world is not governed only by deterministic laws but also by some sort of stochastic one (at least at some times), whose dynamics is captured by Born's Rule. Sometimes, they insisted, the experimental situation "forces" a system to acquire a new property of a sort that it previously did not have. For example, they would say, a system whose wavefunction is a superposition of different position eigenstates, or whose wavefunction is a superposition of different momentum eigenstates, simply *fails* to have any position (or momentum) at all. Still, if one performs a "position measurement" or "moment measurement" on such a system it is required to "jump into" one determinate position or momentum despite not having had one! It could jump into any of several different such determinate eigenstates, and the chance of each outcome is given by Born's Rule. So that part of the dynamics is probabilistic and random. No antecedent physical fact about the system determines which of the various possibilities is realized. God plays dice with the universe.

Einstein is famous for complaining about, and rejecting, the idea that "God plays dice". But from the outset, the indeterminism in the Copenhagen approach was not what really bothered him. What he found completely unacceptable was rather the accompanying "telepathy" or "spooky action-at-a-distance" that the approach demanded. Indeed, in a famous but often bowdlerized statement, Einstein said "It seems hard to sneak a look at God's cards. But that he plays dice and uses 'telepathic' methods (as the present quantum theory requires of him) is something that I cannot believe for a single moment." The real issue for Einstein with Copenhagen was not the "plays dice" part but rather the "uses telepathic methods" part. This was already illustrated in the example of the single radioactive atom and spherical detector.

Suppose that the wavefunction assigned to such a system is complete. According to Schrödinger's equation, the wavefunction of the atom will continuously "leak out" in a spherically symmetric way, and propagate out to the surface of the spherical detector while retaining the spherical symmetry. Accounting for how a spot could form in any one particular place, then, demands breaking the symmetry via a fundamentally stochastic random process: the "collapse of the wavefunction". Bohr's commitment to the completeness of the wavefunction therefore forces him into breaking the Schrödinger evolution in favor of some stochastic evolution: the "collapse". It is exactly at the point of collapse where the spherical symmetry gets broken and Born's Rule replaces Schrödinger. And, famously, the Copenhagen approach places the collapse exactly where a "measurement" occurs.

This skeleton of a theory has several components, all of which Einstein explicitly objected to. One was the indeterminism. A second was placing the indeterminism at the point where a measurement or observation occurs, because there was no sharp characterization of exactly when such a thing occurs. Both of these objections are well known. But there is a third aspect of the theory that Einstein rejected, namely the non-locality of the collapse. Copenhagen collapses had to be both *global* and *instantaneous*, which meant that they effectively occur faster than light. And that runs afoul of Einstein locality.

Let's be entirely clear about how the non-locality manifests in the single atom with spherical detector. The initial wavefunction is spherically symmetric, and the Schrödinger evolution causes it to slowly and continuously radiate out from the atom in a spherically symmetric way. Hence a spherically symmetric wavefunction from the atom reaches the spherical detector, maintaining the symmetry. Of course, the spot will typically not form at the earliest contact of the atomic wavefunction with the screen: the *time* the spot forms is quite variable from experiment to experiment. But at some point, a spot will form in some random place, at some random time, breaking the symmetry. That event will be associated with the collapse of the quantum state. And when the collapse occurs, two different but related things happen.

Most prominently, somehow or other a spot forms on the surface of the detector. But a second consequence of the collapse, beside the *formation* of the spot, is the *disappearance* of the atomic wavefunction *everywhere else*. Once a spot has formed in one location, the chance of a spot forming *anywhere else* goes immediately to zero since it never happens that *two* spots form. For Bohr, the sudden *disappearance* of the quantum state everywhere else but where the spot forms is a real physical change in the state of the world. But the question is: how does a point on the spherical detector at the antipodes of where the spot forms *know that a spot has formed so far away*? According to the Copenhagen understanding, a moment before the spot formed, a moment before the collapse, there was a real physical chance for a spot for form anywhere on the detector, while immediately once the spot formed the physical chance of a spot forming anywhere else became strictly zero. That requires the collapse to be global and instantaneous, and therefore faster than light. If the collapse were not instantaneous, there would be some chance of *two* spots forming, and that never occurs.

Adding insult to injury, in Einstein's eyes, the postulate of this sort of instantaneous telepathy in the single-particle example was obviously easily avoidable. Einstein, like any reasonable person, thought that the appearance of the spot somewhere was the result of *an actual physical particle travelling from the central atom to the detector*. A moment before the spot appeared, the actual particle was very near to the place where the spot would soon appear, and headed that way. In contrast, the Copenhagen account insisted that just before the spot formed *there was no particle in any location at all*. The wavefunction associated with the particle is spherically symmetric and expanded outward, thick enough to be consistent with all sorts of different moments when the spot would form. Since Copenhagen regarded the wavefunction as providing a complete description of the particle, getting a spot in some particular place and at some particular time required a completely different sort of process than the Schrödinger evolution of the wavefunction: it required a *collapse*. One effect of the collapse was to trigger the formation of the spot at a *randomly* selected time and place. Another was to reduce the wavefunction to zero at all *other* places for all *later* times. So collapse of the wavefunction in the Copenhagen approach is exactly the place where both indeterminism ("God plays dice") and non-locality ("telepathy", "spooky action-at-a-distance") is implemented in the theory. Since the *phenomena* did not require either of these peculiar features, Einstein vociferously objected.

His complaints—and the similar complaints of Schrödinger and de Broglie—fell on deaf ears. Einstein could never understand the drift of Copenhages's response to his objections, and eventually he simply gave up trying to convince them. Until 1935.

In that year Einstein, Boris Podolsky, and Nathan Rosen presented a novel example designed to make his point in an even more dramatic fashion.[3] The new set-up employed not one particle but two, each of which could be sent off to a separate lab. The distance between the labs, according to the quantum-mechanical principles, makes no difference to the predictions. Today such distantly separated labs are always run by experimentalists named Alice and Bob.

In the EPR set-up, a pair of particles is prepared in a *maximally entangled* quantum state, and each particle is then dispatched to one of the labs. The mark of entanglement from a purely mathematical perspective is that the wavefunction of each particle cannot be specified without reference to the other. Entangled states exhibit a sort of mathematical *holism*, as if the pair could not be regarded as just the sum of two independent parts even when those parts are unproblematically "situated in different parts of space". The aim of the EPR paper was to argue that that seeming physical interdependence of the particles was just a mathematical illusion, created by the fact that the wavefunction does *not* provide a complete description of the individual system.

In the 1935 paper the entanglement concerned the *positions* and *momenta* of the particles, which were the observable characteristics most commonly discussed in foundational arguments at the time. David Bohm, in his 1951 textbook *Quantum Theory*[4], changed the example to use the characteristic called "spin", and we will follow Bohm here. The "spin" of a spin-½ particle can be "measured" by passing it through an inhomogeneous magnetic field, using a Stern-Gerlach apparatus. In each individual run of the experiment at spot forms on the screen either at a fixed distance above or a fixed distance below the midline (see Figure 1). If the spot forms in the direction of the pointy magnet, the result is called "spin up" and if in the direction of the flat magnet "spin down". In this experimental set-up the possible results are "quantized" or discrete: each outcome is either spin-up or spin-down. (The corresponding classical set of possible outcomes is shown in Figure 1: in that theory the spot could form at any location between the two maximal deflections. The eye-shape forms because as the particle drifts to one side or the other the inhomogeneity of the magnetic field is reduced, so off to the sides the particles are not deflected) The magnet can be oriented in any direction, so there are in principle an infinite number of "spin measurements" that can be made. But for our purposes we need only consider two possible directions, which are at right angles to each other. We will call these the "x-direction" and the "z-direction".

---

[3] "Can Quantum Mechanical Description of Reality Be Considered Complete?", *Physical Review* **47**, 777-80.
[4] *Quantum Theory*, Prentice-Hall, 1951.

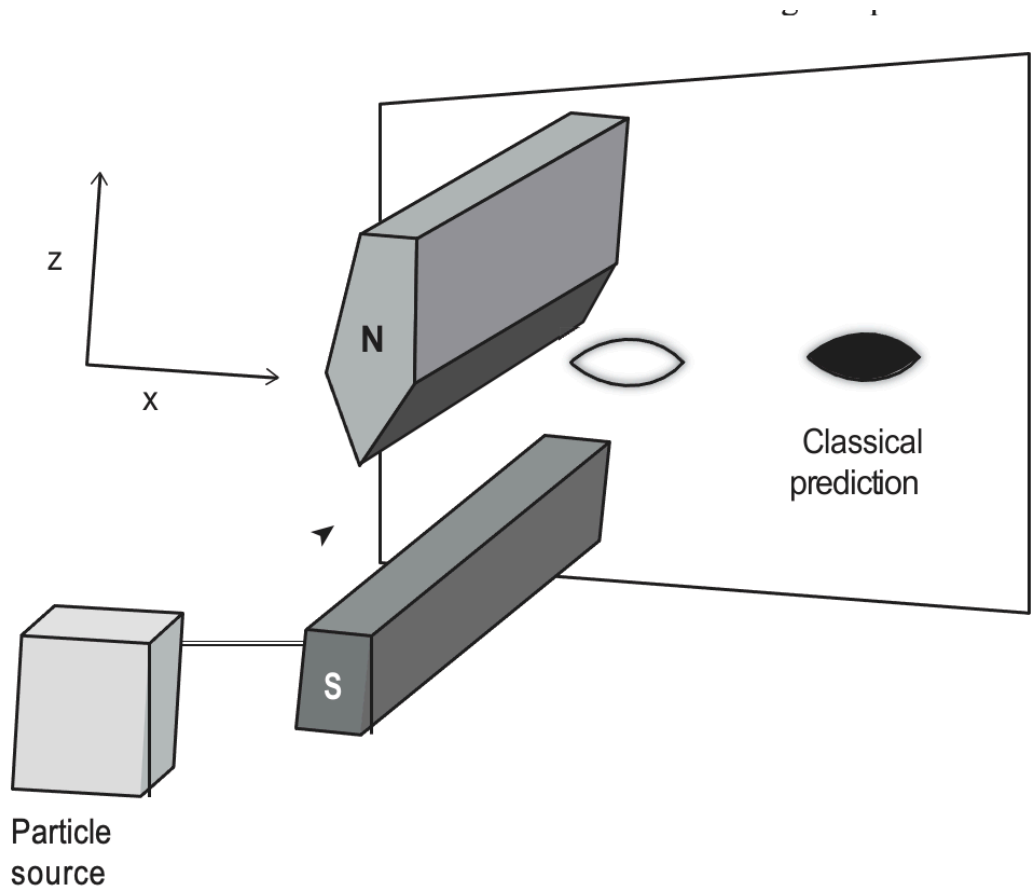

Figure 1: The Stern-Gerlach Apparatus

The EPR argument requires a pair of Stern-Gerlach devices, one in each of two spatially separated labs. The labs should be so far apart that no signal transmitted at light speed could get from one to the other in time to carry information about how the experiment in that lab came out. Schematically the situation is as depicted in Figure 2.

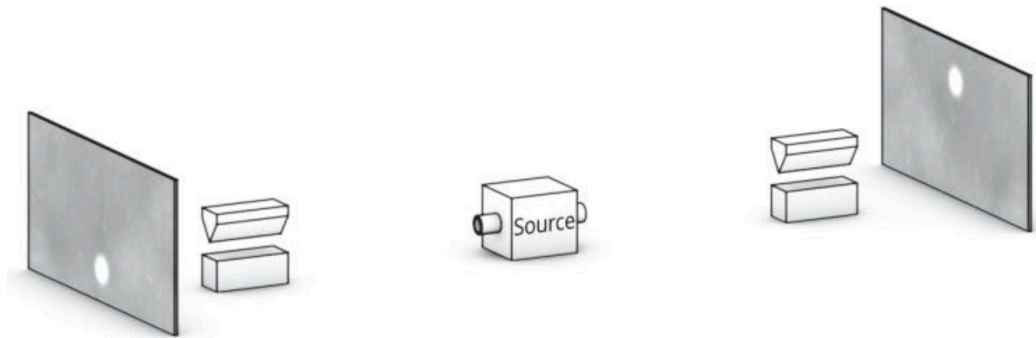

Figure 2: The EPR Set-Up

The EPR experiment starts by preparing the pair of particles in the singlet state, an entangled state which may be written as $\frac{1}{\sqrt{2}}|Z\uparrow>_A|Z\downarrow>_B - \frac{1}{\sqrt{2}}|Z\downarrow>_A|Z\uparrow>_B$ (or alternatively, for example, $\frac{1}{\sqrt{2}}|X\uparrow>_A|X\downarrow>_B - \frac{1}{\sqrt{2}}|X\downarrow>_A|X\uparrow>_B$, which is the same quantum state). The subscript A indicates the particle that is sent to Alice and B the particle that is sent to Bob. What is curious about this quantum state is that it allows for *perfect predictions with certainty* of a sort in this experimental condition. In particular, the standard quantum predictive algorithm asserts that in *every* experimental run Alice and Bob will get *different* outcomes. If Alice records a Z-spin up outcome then Bob with record Z-spin down, and vice versa. (And similarly if they both orient their magnets in the X-direction or in any other direction.) The predictive algorithm *does not* predict which of these two outcomes will occur. In fact, on that point it is maximally uncertain: each outcome has a 50% chance. Nonetheless, that complete uncertainty about the specific outcomes is paired with complete certainly that the outcomes—whatever they are—will be different.

One can just read this prediction off of the singlet state: since the state is expressed entirely in terms of Z-spin, one need only square the weighting factor $\frac{1}{\sqrt{2}}$ to make the prediction. The chance that the pair of particles will behave as a pair in the state $|Z\uparrow>_A|Z\downarrow>_B$, yielding Alice an up outcome and Bob a down outcome, is ½. And similarly for $|Z\downarrow>_A|Z\uparrow>_B$.

The EPR argument makes use of this perfect correlation by reference to the EPR "criterion of reality". A "criterion" is not in this case offered as a *definition*: Einstein, Podolsky, and Rosen do not pretend to offer *necessary and sufficient conditions* for the existence of some element of physical reality. Rather they offer only a sufficient condition, but one they regard as completely unproblematic. Here is the definition from the paper:

> If, without in any way disturbing a system, we can predict with certainty (i.e., with probability equal to unity) the value of a physical quantity, then there exists an element of **reality** corresponding to that quantity.

Application of the criterion requires that two conditions be met: first, someone must somehow be able to get into a position to predict *with certainty* what the outcome of some experiment performed on a target system will be; and second the means of getting into that position must not in any way *disturb* or *change the physical state* of the target system. If that is possible, the reasoning goes, then there must be *some* physical characteristic of the system that is pre-determining it to act that way. Otherwise, how could the perfectly reliable and accurate prediction about how it will behave possibly always be correct? It is hard to see any objection that can be raised to the Criterion of Reality.

Still, the application of the Criterion to any particular case requires accepting that the method by which the predictor comes to make the prediction does not, in fact, disturb or change the target system in any way. (Of course, the method will certainly change the predictor's *beliefs about* or *knowledge of* or *information concerning* the target system, but that is neither here nor there as far as the Criterion goes). So the central question is how one can come to be so confident that the method of prediction does not disturb the target system. And EPR's solution to that question is straightforward and elegant: one comes to be sure that nothing Bob does in his lab can disturb or change or alter the particle in Alice's lab by separating his lab from hers by an arbitrarily great distance. If anything Bob does in that circumstance physically disturbs Alice's particle, or anything she does physically disturbs his, that would require "telepathy" or "spooky action-at-a-distance". Furthermore, if the timing of when Bob makes his prediction and Alice does her experiment is arranged properly, then any such telepathy or action-at-a-distance would have to operate *faster than light*. All of these things which would have to happen in order that Bob's method of prediction disturb Alice's particle are so bizarre and unheard of in classical physics (including the theories of Relativity) that EPR just take for granted that no one would dare claim Bob's procedure disturbs Alices particle. In which case the Criterion would apply.

Bob's method of predicting Alice's outcome is trivial: he simply checks the Z-spin of *his own* particle in *his own* lab. Since the quantum-mechanical entangled state (which they all know was prepared) predicts the perfect anti-correlation of the particles, as soon as Bob gets his result he is in a position to accurately predict with certainty Alice's result: it must be the opposite one. So long as one accepts that Bob's experimental procedure does not disturb the physical state of Alice's particle the EPR argument goes off without a hitch: application of the Criterion implies that Alice's particle must have some element of reality in it that determines the outcome of *her* experiment. The only way to deny to conclusion is to assert that Bob's experiment does indeed disturb Alice's particle…..by telepathy.

Since EPR assume that not even Bohr would embrace such telepathy, they consider their argument decisive. What it proves, in their eyes, is that the quantum-mechanical entangled state $\frac{1}{\sqrt{2}}|X \uparrow>_A|X \downarrow>_B - \frac{1}{\sqrt{2}}|X \downarrow>_A|X \uparrow>_B$ must be *incomplete*. The argument is trivial: since they have shown (if one rejects superluminal telepathy) that Alice's particle contains an element of physical reality that predetermines the outcome of her experiment, and since the entangled state $\frac{1}{\sqrt{2}}|X \uparrow>_A|X \downarrow>_B - \frac{1}{\sqrt{2}}|X \downarrow>_A|X \uparrow>_B$ does not represent any such predetermining feature, then $\frac{1}{\sqrt{2}}|X \uparrow>_A|X \downarrow>_B - \frac{1}{\sqrt{2}}|X \downarrow>_A|X \uparrow>_B$ must not tell the whole story about Alice's particle. It must be incomplete since it leaves out an element of physical reality that her particle has.

One can arrive at the same conclusion by considering two runs of the experiment which yield different outcomes. Suppose that on the first run the outcome is Alice gets Z-up and Bob gets Z-down, while on the second Alice gets Z-down and Bob gets Z-up. According to the quantum-mechanical formalism, the *complete* physical state of the pair of particles, and of the laboratory equipment, was *exactly the same* at the outset of the experiment. Nonetheless, the two runs yielded different outcomes. Therefore the fundamental physical dynamics would have to be *indeterministic*: the same initial state can evolve into different final states. Or, to put in in Einstein's pungent phrase, according to such a theory "God play dice with the universe".

But note well: the objection that EPR raise against this account is not the dice-playing *per se*. If the physical world is fundamentally indeterministic, then all well and good as far as this argument goes. But if a physical interaction is fundamentally indeterministic, then *no one should be able to predict with perfect accuracy how it will come out*! But in this situation, Bob *can* predict—on the basis of *his* outcome—how Alice's experiment will (or did) come out. If neither of their experiments has any *physical, causal* influence on the other then how could that be? If, Alice's outcome, for example, is truly indeterministic and unpredictable, and it further has no effect on Bob's particle, then how is it that Bob's particle always does the opposite thing? To put is only slightly anthropocentrically, how does Bob's particle *know what to do* in order that it result be the opposite of Alice's?

As John Bell remarks in "Bertlmann's Socks and the Nature of Reality"[5], EPR did not consider themselves to be proposing a *paradox* but rather just making a fairly trivial *observation*. Bell considers another perfect anti-correlation, that between the two socks that Reinhold Bertlmann wore each day, which were *always* of different colors. There was no paradox or mystery about that perfect anti-correlation: Bertlmann just intentionally always chose socks of different colors when he got dressed each morning. But note: if we take Bertlmann getting dressed as the analog of the creation of the particles in the EPR set-up, then their reactions to the experimental conditions they encounter in Alice's and Bob's labs *must be predetermined from the start*. If one insists that they were not—if one insists (as Bohr would)—that neither particle *has* a definite Z-spin until it is "measured", that the experiments done in the labs somehow *force* the particles to choose which Z-spin they will display—then one certainly has an extraordinary hypothesis. It is, of course, somewhat extraordinary that the experimental conditions could "force" a particle to make a *random* choice. But even more remarkable—much, much more remarkable—is that the two particles somehow manage to make *different* choices! As Bell puts it:

> It is in the context of ideas like these that one must envisage the discussion of the Einstein-Podolsky-Rosen correlations. Then it is a little less unintelligible that the EPR paper caused such a fuss, and that the dust has not settled even now. It is as if we had come to deny the reality of Berlmann's socks, or at least of their colors, when not looked at. And as if a child had asked: How come they always choose different colors when they *are* looked at? How does the second sock know what the first has done?
>
> Paradox indeed! But for the others, not for EPR. EPR did not use the word 'paradox'. They were with the man in the street in this business. For them, these correlations simply showed that the quantum theorists had been hasty in dismissing the reality of the microscopic world. In particular Jordan had been wrong in supposing that nothing was real or fixed in that world before observation. For after observing only one particle the result of subsequently observing the other (possibly at a very remote place) is immediately predictable. Could it be that the first observation somehow fixes what was unfixed, or makes real what was unreal, not only for the near particle but also for the remote one? For EPR that would be an unthinkable "spooky action at a distance". To avoid such action at a distance they have to attribute, to the space-time regions in question, *real* properties in advance of observation, correlated properties, which *predetermine* the outcomes of these particular observations. Since these real properties, fixed in advance of observation, are not contained in quantum formalism, that formalism for EPR is *incomplete*. It may be correct as far as it goes, but the usual quantum formalism cannot be the whole story.[6]

---

[5] J.S. Bell, *Speakable and Unspeakable in Quantum Mechanics*, Second Edition, Cambridge University Press, 2008, Chapter 16.

Bell has summed the whole thing up. The EPR set-up leaves us with a choice, one side of which was completely unpalatable to Einstein. Either the quantum formalism is incomplete, leaving out specification of the real properties that predetermine the outcomes for each particle in the EPR situation, or else one must appeal to some non-local telepathy or spooky action at a distance that violates the locality condition described in the letter to Born above. For Einstein, the choice was, as we would say, a no-brainer. For there was absolutely nothing at the time supporting the completeness of the quantum formalism except for Copenhagen's blunt insistence that it must be so.

And so things stood until Bell.

Before getting to Bell's Theorem (and its variants), it is worthwhile noting how powerful the EPR argument is in the context of experiments done on the singlet state. We have seen that the only way that a *local* physics can predict or account for the perfect EPR correlations between the outcomes of Z-spin experiments is to attribute a definite Z-spin to each particle even *before* it enters Alice's or Bob's lab. These values must be assigned to the particles at the source if no telepathy is to be appealed to. But we have already seen that the expression of the sinlget state in terms of Z-spin is identical to its expression in terms of X-spin, so the same argument can be made: given the verifiable perfect anti-correlation whenever both Alice and Bob set their devices in the X-direction, if physics is local then each particle must be assigned an X-spin at the source (with the two particles getting different assignments). And indeed the singlet state takes the same form when expressed in terms of spin in *any* direction. So long as Alice and Bob set their devices in the same direction they will always get opposite outcomes, so in a local theory they spin in *every* direction must be assigned at the source.

This is a particular embarrassment for those maintaining that the quantum wavefunction is complete because no wavefunction can predict the outcome of a spin experiment in more than one direction. So if one wants to continue with local physics, one *must* use something other than quantum-mechanical wavefunctions to describe particles.

---

[6] Bell *ibid* p. 143.

The question that occurred to Bell, regarding the quantum-mechanical predictions for the singlet state, is what happens if Alice and Bob set their respective apparatuses not in the same direction, nor in orthogonal directions, but rather with some intermediate angle between them. The quantum formalism makes precise predictions about the degree of correlation between Alice's and Bob's outcomes in all these cases, and what Bell proved is that no local theory can reproduce those predictions. Bell's argument is not difficult—it amounts to a few lines of simple algebra—but a later discovery by Daniel Greenberger, Michael Horne, and Anton Zeilinger (the GHZ example) makes Bell's point in a particularly sharp and clear way. So we will bypass Bell's original argument in favor of the GHZ set-up.

Just as EPR went from consideration of single particles to consideration of entangled pairs, the GHZ argument passes to consideration of entangled triplets. The experimental situation is depicted in Figure 3. In addition to Alice and Bob, a third experimenter (Charlie) is given his own lab. Just as in the EPR example, particles will be created at the source in a specific entangled quantum-mechanical state and then sent off the three labs, which can be situated as far apart from one another as one likes. In particular, they can be so far apart that no light signal announcing the outcome of the experiment done in one lab can reach any of the other labs in time to have any influence on their experiments. If the physics of the universe is local, then the three labs will be causally isolated from one another.

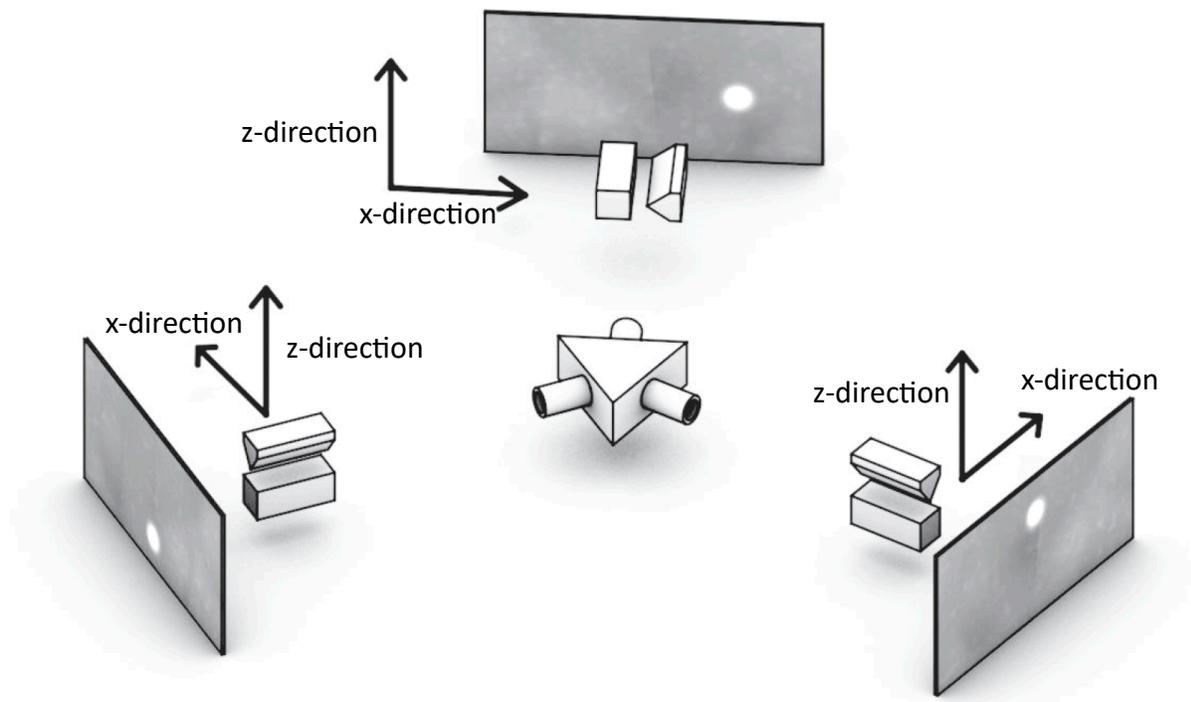

Figure 3: The GHZ Experiment

In the GHZ set-up Alice, Bob, and Charlie each can make a choice between two possible experimental settings of their magnets, which we will call the "Z-direction" and the "X-direction". The experimenters determine the direction on each run in whatever way they like. It is often said that they can use their "free will" to determine the setting, but that locution is both misleading and extremely impractical. The desideratum is that the setting be determined in a way that is *statistically independent* of the states of the incoming particles, that is, the settings are not correlated with, and carry no information about, what sorts of particles will enter the lab. One way to ensure this statistical independence is as far from "free choice" as possible: Bell mentions the possibility of having the setting determined by the parity of the digits of some input variable, such as pi, beginning at the millionth digit. Such a mechanism clearly involves no one's "free will", but the choices will be made randomly for all practical purposes. That is, the sequence of Z-settings and X-settings will pass every statistical test for a random sequence, showing no predictable patterns. It is, in Bell's phrase, "sufficiently free for the purposes at hand".

The same randomness—that is, statistical independence of the apparatus settings from the states of the incoming particles—can also be achieved by mechanical means. Much thought has gone into this problem for the purpose of randomly choosing winning lottery numbers, and the result is "physical randomizing devices". Bell again:

> Consider the extreme case of a 'random' generator which is in fact perfectly deterministic in nature—and, for simplicity, perfectly isolated. In such a device the complete final state perfectly determines the complete initial state—nothing is forgotten. And yet for many purposes, such a device is precisely a 'forgetting machine'. A particular output is the result of combining so many factors, of such a lengthy and complicated dynamical chain, that it is quite extraordinarily sensitive to minute variations of any one of the many initial conditions. It is the familiar paradox of classical statistical mechanics that such exquisite sensitivity to initial conditions is practically equivalent to complete forgetfulness of them. To illustrate the point, suppose that the choice between two possible outputs, corresponding to *a* and *a'* [e.g. to setting the magnet in the Z-direction or the X-direction], depended on the oddness or evenness of the digit in the millionth place of some input variable. Then fixing *a* or *a'* indeed fixes something about the input—i.e. whether the millionth digit is odd or even. But this peculiar piece of information is unlikely to be the vital piece for any distinctly different purpose, i.e., it is otherwise rather useless. With a physical shuffling machine, we are unable to perform the analysis to the point of saying just what particular feature of the input is remembered in the output. But we can quite reasonably assume that it is not relevant for other purposes. In this sense the output of such a device is indeed a sufficiently free variable for the purposes at hand. For this purpose the assumption (1) [of his theorem] follows.[7]

Bell's discussion of the "free choice" of the experimental settings in "Free variables and local causality" (Chapter 12 of *Speakable and Unspeakable in Quantum Mechanics*) is the *locus*

---

[7] Bell, ibid, pp. 102-103.

*classicus* for discussion of this topic. Note that none of Bell's "physical randomizing devices" appeals to a speck of "free will" on the part of the experimenters, and further that that the particular devices he mentions are stipulated to be *deterministic* in nature. A tremendous amount of wasted ink has been spilled on this topic by people who seem to think that Bell' Theorem only follows if one grants some sort of contentious claims about human free will. Nothing could be farther from the truth, and the people spilling the ink have evidently not read "Free Variables and Local Causality" or not read it carefully.

We will simply grant that in the GHZ set-up Alice, Bob and Charlie all use physical randomizing devices to set their apparatuses on each run, and that therefore those settings can be taken to be statistically independent of the physical state of the incoming particles.

In the GHZ experiment Alice, Bob, and Charlie each makes (or has made for them) a binary choice of experimental condition on each run of the experiment. Therefore, on each run there are eight possible global experimental conditions, which can be conveniently indicated as $Z_A Z_B Z_C$, $X_A Z_B Z_C$, $Z_A X_B Z_C$, $Z_A Z_B X_C$, $X_A X_B Z_C$, $X_A Z_B X_C$, $Z_A X_B X_C$, and $X_A X_B X_C$. The usual quantum algorithm makes statistical predictions about the outcomes for each of the eight possible situations, but for our purposes only four of them matter: $Z_A Z_B Z_C$, $X_A X_B Z_C$, $X_A Z_B X_C$, and $Z_A X_B X_C$. Data for the other runs play no role in the argument.

In each of the four remaining global experimental arrangements the quantum formalism makes some predictions *with complete certainty*, just as it did in the EPR situation. As with the EPR case, the predictions are not about exactly what result each experimentalist will get, but rather about correlations in their their collective data. In particular:

> Prediction 1: When the global experimental arrangement $Z_A Z_B Z_C$ happens (by chance) to be chosen, then there will certainly be recorded an *odd* number of "up" outcomes. It could be 1 and it could be 3 (and the quantum algorithm will assign precise probabilities to each), but it will certainly, in every case, be odd.

> Prediction 2: In any of the other three possible global experimental arrangements, viz. $X_A X_B Z_C$, $X_A Z_B X_C$, and $Z_A X_B X_C$, there will certainly be an *even* number of "up" outcomes among the three. It could be 2 and it could be 0, but it will never be 1 or 3.

All one needs to know about the GHZ experiment is that there exists an entangled quantum spin state of the three particles that makes these predictions. In the sequel, we assume these predictions are accurate. They have been tested in the laboratory and found to be so.

Note that each of the predictions above is an example of an EPR correlation in the sense that they allow for the application of the EPR Criterion of Reality (assuming the physics is local in Einstein's sense). For example, suppose the global experimental situation happens to be $Z_A Z_B Z_C$. Then from the known outcome of any two of the labs the outcome in the third lab can be predicted with certainty. Given the outcome of the two, only one outcome for the third will be compatible with the predictions. So *on the assumption that the experimental operations and outcomes in each of the labs has no physical influence on—does not 'disturb'—the physical state in the other labs, the EPR argument applies and one can conclude that the outcomes of each of the three experiments must be predetermined at the source.* For if there were any fundamentally indeterministic or "chancy" aspect to the dynamics leading to the outcome in one lab, in a local theory there would be no way for that outcome to influence the outcomes in the other labs and hence no way for the outcomes in the other labs to adjust themselves so that the prediction will be certainly fulfilled.

So by the very argument that EPR used with respect to their set-up, we get the result that in any physics with a local dynamics in Einstein's sense (no "telepathy" or "spooky action at a distance"), the outcomes of *all three* experiments must be predetermined at the source, and indeed must be predetermined for *both* of the possible experimental situations that Alice, Bob and Charlie may choose since there is no way that that choice could be known or have any influence on the particles at the time of their creation. In short, in a local theory how each of the three particles would react to a Z-oriented magnet and to an X-oriented magnet must already be determined at the source, and the dynamics that connects the state of the particles at their creation to the later outcome of the experiments done on them must be deterministic. That is what locality demands in this case.

But now comes the rub. *It is mathematically impossible to predetermine the results of all six possible local experimental conditions so as to guarantee Prediction 1 and Prediction 2.* It simply cannot be done. *No* local theory, no theory with a local dynamics in Einstein's sense, can make the same predictions in this case as quantum theory does—and as is confirmed to occur in the lab.

The proof of the mathematical fact that it is impossible to predetermine the outcomes in each lab in any local theory—i.e. in any there where the outcome in one of the labs is uninfluenced by which experiment was carried out in any other lab and what outcome it had—is spectacularly simple and compelling. It helps to use the following diagram (Figure 4) of the situation, originally introduced (to my knowledge) by David Mermin:

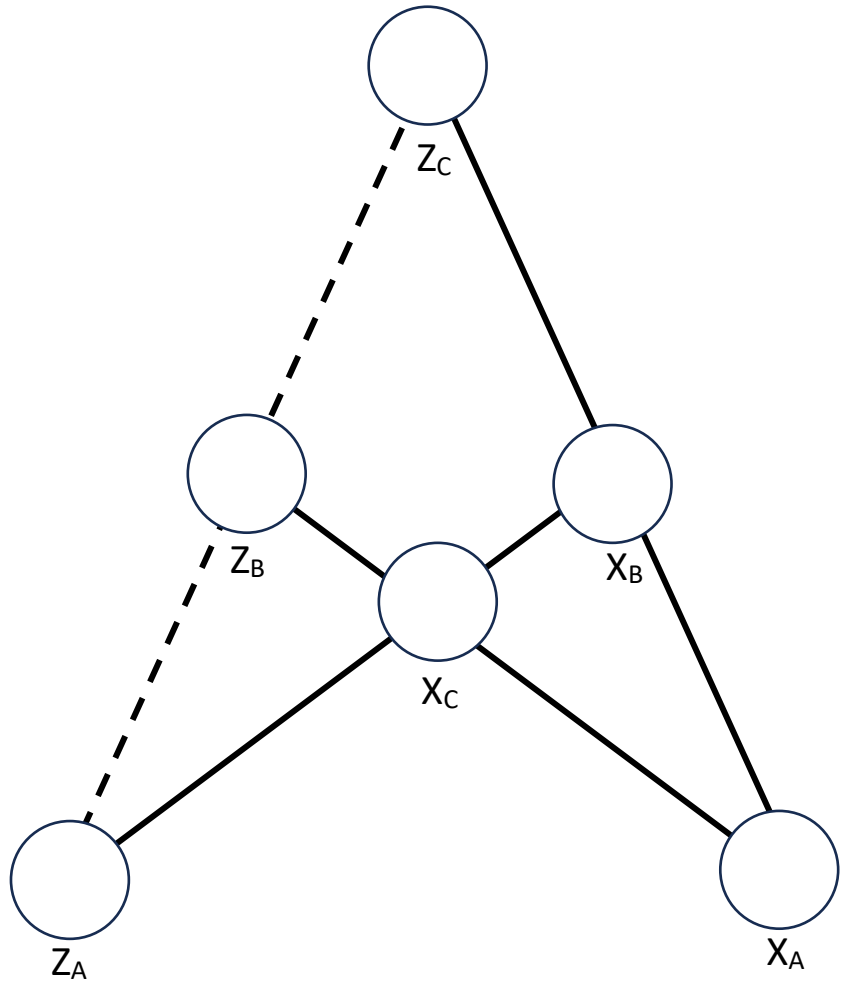

Figure 4: The GHZ Situation Graphically

Note that in the diagram the experimental situation corresponding to Prediction 1—all Z-orientations—is represented by the dashed line while the other three situations corresponding to Prediction 2 are represented by the three solid lines. So one's task is simple: write either the letter "U" or the letter "D" in each of the six circles in such a way that all four of the predictions will be fulfilled, that is, so that there are an odd number of "U"s along the dashed line (in case that experimental arrangement is randomly chosen) and an even number along each of the three solid lines (in case any of those three experimental arrangements is randomly chosen).

After a little experimentation one will quickly come to suspect that it can't be done. The proof of this that Mermin offered is delightfully elegant. It is a proof by *reductio*. Suppose the task could be accomplished, so there is either a "U" or a "D" in each of the six circles. Imagine them as disks with the letters printed on them. Now pick up the three disks that lie along the dotted line and throw them in a hat. One has just put an odd number of "U"s into the hat. Now pick up the three disks that lie along each of the solid lines that throw then into the hat. If the conditions of the predictions have been fulfilled, each of those sets of three contains an even number of "U"s. So having put an odd followed by three sets of even numbers of "U"s in the hat, there must now be a total of odd number of "U"s in the hat. But each disk was picked up twice, since each circle lies on the intersection of two lines. So no matter how the disks were lettered, there cannot possibly be an odd number of "U"s in the hat. Ergo the task in impossible. QED.

In short, *no possible dynamically or causally local physical theory can reproduce the quantum-mechanical predictions*. And, much more importantly, since the quantum-mechanical predictions are in fact true, *the physics of the actual world—whatever it is—cannot possibly be local*. The non-locality Einstein objected to in quantum theory cannot be somehow removed by changing to another sort of theory. The non-locality is here to stay because the predictions of quantum theory in situations like this *are accurate*. That is not a theoretical claim but an experimental one: it has been confirmed in the lab.

Now we reach the end of the story. The confirmed predictions of quantum theory are not merely hard to reconcile with Einstein's vision of a completely local physics—a physics in which both the ontology and the laws are defined and can be checked locally and nothing propagates faster than light—they are flatly incompatible with Einstein's vision. And since General Relativity was the final expression of Einstein's question for locality, they are flatly incompatible the General (or Special) Relativistic account of space-time structure and the way dynamical laws are articulated in terms of the space-time structure. Between quantum theory and Relativity something has to give, and the thing that has to give is Relativity.

The present-day problem is that physicists in general have not accepted this conclusion. In large part, that is because most physicists have not understood what Bell proved. A very common take on Bell's Theorem is that it rules out only "hidden variables theories" or only "deterministic theories". Both of these claims are radically false.

Bell's Theorem cannot have ruled out deterministic theories because violations of his inequality are predicted and accounted for by the pilot wave theory (Bohmian Mechanics), and that theory is completely deterministic. So determinism need not be abandoned. Einstein himself was aware of the theory from de Broglie, and his objection to it had nothing to do with determinism or indeterminism—in fact, Einstein was inclined to expect the correct theory to be deterministic *because he expected it to be local*. Einstein's complaint about the theory was rather its manifest *non-locality*. The "spooky action at a distance" in the pilot wave approach is easy to see in the equations, and it is for that reason that Einstein rejected theory. What Bell proved is that the non-locality is not an avoidable or option feature of any empirically adequate theory.

What about "hidden variables"? If all Bell did was to somehow rule out "hidden variables", then most physicists could rightfully yawn: they were never attracted to "hidden variables" in the first place. But this understanding of Bell's work again makes no contact with what he actually did. Indeed, the concept of a "hidden variable" nowhere appears in his proof and nowhere *could* appear in his proof. For the concept only makes sense in the context of quantum theory, and Bell's proof nowhere mentions or adverts to quantum theory. Rather, what Bell shows is that *any* theory, built on *any* mathematical or conceptual foundation must incorporate some form of non-locality if it is to make certain predictions. It is true that those predictions happen to be made by quantum theory, but that is immaterial to the proof. And, as we have insisted, the really important thing about those predictions is not that they can be derived from the quantum-mechanical predictive formalism but that they actually are observed in the lab.

What Bell rules out—given those empirical results—is neither determinism nor "hidden variables". It can't be, because Bohmian mechanics predicts violations of Bell's Inequality and it is both deterministic and does postulate "hidden variables". What Bell rules out, purely and simply, is locality. And since that form of locality, advocated by Einstein and used by Bell in his proof, is built into the foundations of Relativity what Bell ultimately rules out is Relativity. What sort of space-time structure should be used in place of the space-time of Special Relativity or the space-times of General Relativity is not at all clear, and should be a question given the highest level of attention by anyone seeking to "reconcile" quantum theory with Relativity.